\documentclass[sigconf]{acmart}

\AtBeginDocument{%
  \providecommand\BibTeX{{%
    \normalfont B\kern-0.5em{\scshape i\kern-0.25em b}\kern-0.8em\TeX}}}

\acmConference[Vinci '24]{Siggraph}{December 11--13, 2024}{Hsinchu, Taiwan}
\acmBooktitle{Vinci '24, December 11--13, 2024, Hsinchu, Taiwan}

\usepackage{pdfpages}
\begin{document}

\title{ Visions of Destruction: Exploring a Potential of Generative AI in Interactive Art }

\author{Mar Canet Sola }
\authornotemark[1]
\email{mar.canet@gmail.com}
\orcid{0000-0001-5986-3239}
\affiliation{%
  \institution{BFM, Tallinn University, Estonia }
  \country{}
}
\affiliation{%
  \institution{ Academy of Media Art Cologne, Germany}
  \country{}
}

\author{Varvara Guljajeva }
\authornote{Both authors contributed equally}
\email{varvarag@gmail.com}
\orcid{0000-0002-0261-3121}
\affiliation{%
  \institution{ Academy of Media Art Cologne, Germany}
  \country{}
}

\begin{teaserfigure}
  \centering
  \includegraphics[width=\linewidth]{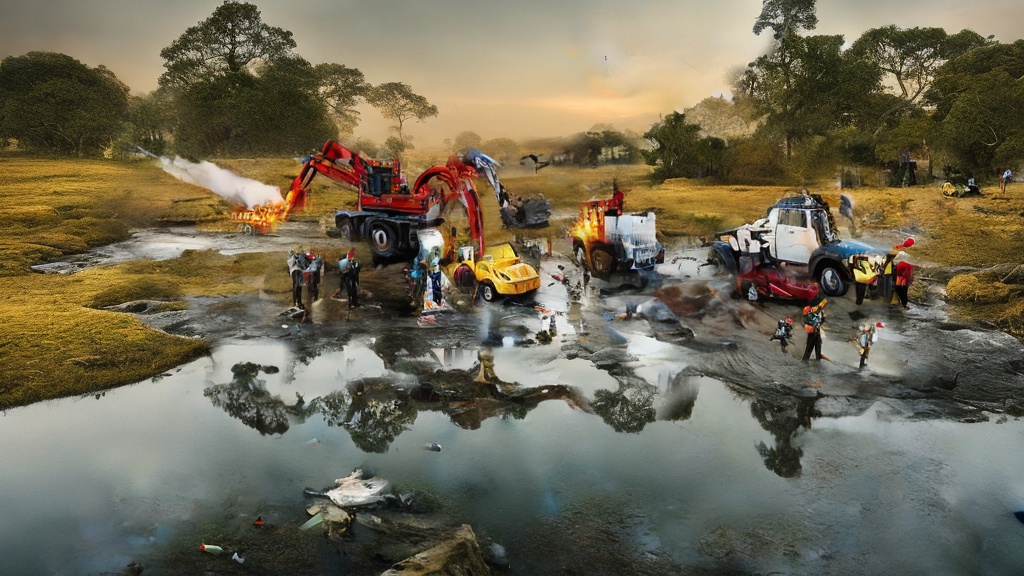}
    \caption{ 
    A visual composition generated as a transformed landscape view through audience interaction via gaze tracking. "Visions of Destruction" (2023) by Varvara \& Mar.}
  \Description{ }
  \label{fig:cvpr_output}
\end{teaserfigure}

\begin{abstract}

This paper explores the potential of generative AI within interactive art, employing a practice-based research approach. It presents the interactive artwork "Visions of Destruction" as a detailed case study, highlighting its innovative use of generative AI to create a dynamic, audience-responsive experience. This artwork applies gaze-based interaction to dynamically alter digital landscapes, symbolizing the impact of human activities on the environment by generating contemporary collages created with AI, trained on data about human damage to nature, and guided by audience interaction. The transformation of pristine natural scenes into human-made and industrialized landscapes through viewer interaction serves as a stark reminder of environmental degradation. The paper thoroughly explores the technical challenges and artistic innovations involved in creating such an interactive art installation, emphasizing the potential of generative AI to revolutionize artistic expression, audience engagement, and especially the opportunities for the interactive art field. It offers insights into the conceptual framework behind the artwork, aiming to evoke a deeper understanding and reflection on the Anthropocene era and human-induced climate change. This study contributes significantly to the field of creative AI and interactive art, blending technology and environmental consciousness in a compelling, thought-provoking manner.

\end{abstract}




\maketitle

\begin{figure}[htbp]
  \centering
  \includegraphics[width=\linewidth]{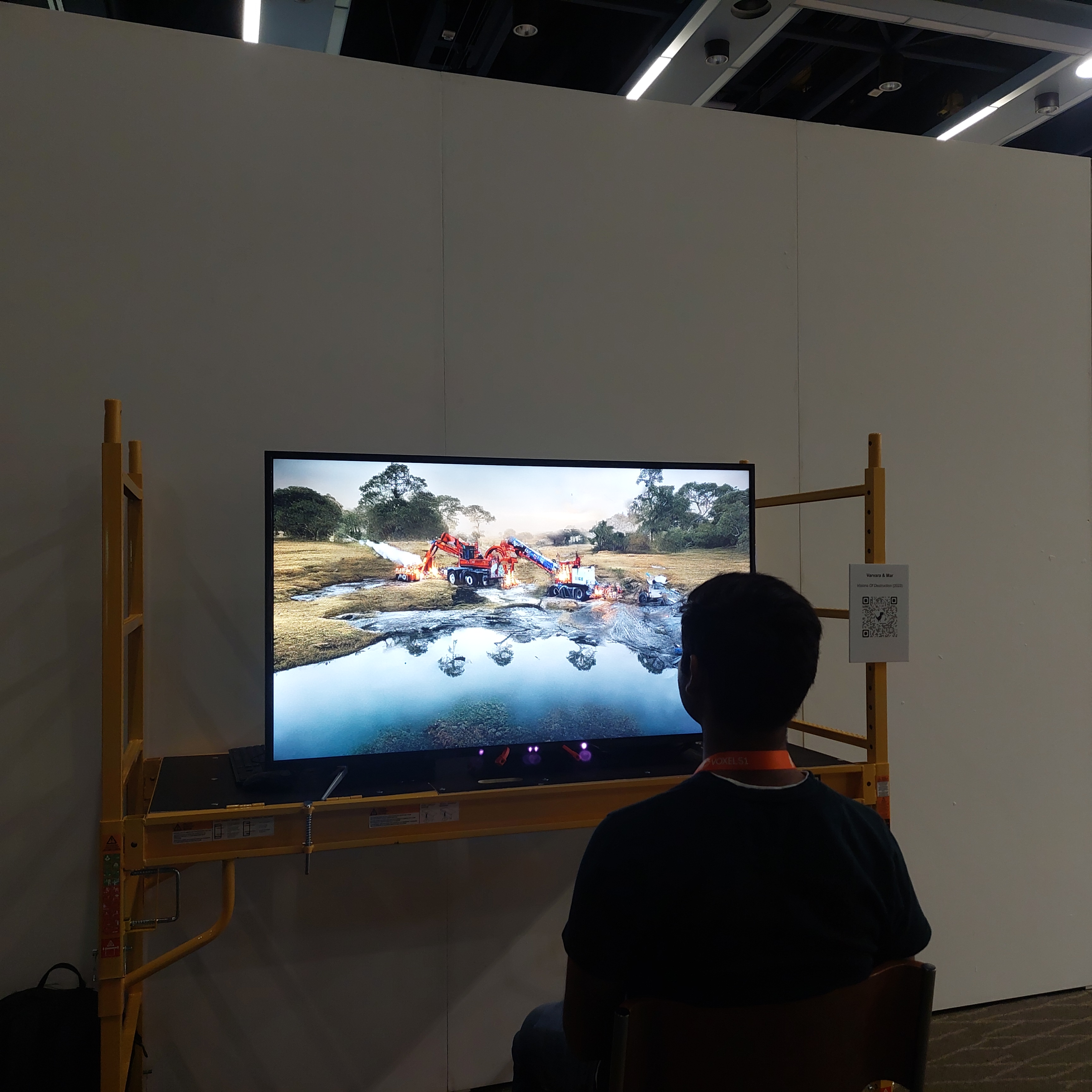}
  \caption{A participant interacting with artwork "Visions of Destruction" (2023) by Varvara \& Mar.}
  \Description{A participant interacting with artwork "Visions of Destruction" (2023) by Varvara \& Mar.}
  \label{fig:cvpr_interactions}
\end{figure}

\section{Introduction}

The rapid advancements in deep learning (DL) have significantly broadened the capabilities of generative AI, which radically transformed the creative methodologies used by creators in conceptualizing ideas and transitioning them into production \cite{epstein2023art}. GANs, first introduced in 2014 \cite{goodfellow2014generative}, initially produced outputs with limited diversity, resembling their training datasets.
Our previous practice-based research confirms this statement and also underlines that not harmonious datasets produce rather abstract results \cite{guljajeva2022postcard}.

This limitation was partly due to the use of smaller dataset sizes. For instance, StyleGAN3 \cite{karras2021alias} was released in late 2021 with three pre-trained datasets, each featuring different content types: the FFHQ dataset includes 70,000 high-quality images for human faces, AFHQv2 \footnote{https://huggingface.co/datasets/huggan/AFHQv2} offers around 12,200 diverse animal face images, and MetFaces \footnote{https://github.com/NVlabs/metfaces-dataset}, designed explicitly for artistic face generation, contains a collection of only 1,336 images. These datasets are still considerably smaller compared to those used by diffusion models. For example, Stable Diffusion released in 2022 was trained on the extensive LAION-5B dataset \footnote{\url{https://laion.ai/laion-5b-a-new-era-of-open-large-scale-multi-modal-datasets/}}, which comprises over 5 billion images with mixed content type, showcasing the evolution in dataset scale for generative models. However, advancements in GAN architectures, such as StyleGAN-XL \cite{sauer2022stylegan} and StyleGAN-T \cite{sauer2023stylegan}, demonstrate significant improvements and show that the model also can work with large diverse datasets like ImageNet\footnote{https://image-net.org/} of 14 million. Moreover, advancements in methods for navigating the latent space of AI models, such as image-to-text techniques \cite{mansimov2015generating,reed2016generative} and image-to-image inpainting methods \cite{rombach2022high} using diffusion models, bring new opportunities for interactive art. These innovations enable more precise and diverse manipulations within the latent space for image generation, enhancing the creative potential and application scope of generative models.

The study by Miyazaki et al. explores public perception towards generative AI, particularly in creative professions \cite{miyazaki2023public}. According to their findings, artists and other creative professionals often harbor negative sentiments towards generative AI technologies \cite{miyazaki2023public}. This apprehension predominantly stems from concerns regarding the training datasets used by these AI systems, which frequently contain copyrighted material from artists \cite{ho2024midjourney} and other creatives. Such issues raise important ethical and legal questions about the sourcing and use of data in AI development, along with potential harms to artists \cite{jiang2023ai}. However, it is noteworthy that many artists have embraced DL tools within their creative workflows for content generation \cite{cetinic2022understanding}. Concurrently, the concept of collaborative co-creation with AI tools is making inroads into educational settings \cite{fiebrink2019machine}.

This article aims to shed light on the exciting opportunities that generative AI presents within the field of interactive art, especially for alternative forms of real-time content generation guided by the users' interactions (see Figures \ref{fig:cvpr_interactions}, \ref{fig:cvpr_output}). 

\subsection{Generative AI and Interactivity}

Advancements in AI, particularly in the domain of computer vision, have significantly revolutionized the field of interactive art. This genre, which dynamically responds to the participant's movements and actions, traditionally relied on interfaces comprising cameras, lights and crafting custom-made software. However, the integration of DL techniques has simplified and enhanced the tracking mechanisms in these installations. Previously, these systems required complex and expensive setups with precise calibrations, often utilizing specialized equipment like infrared cameras with controlled lighting, thermal cameras, or depth cameras to accurately capture body silhouettes or detect specific body parts such as the face, hands, or eyes. DL models trained on large datasets have streamlined these processes, simplifying the setups in many cases to standard RGB webcams, reducing the complexity of calibration and improving the tracking accuracy. This evolution makes the technology more accessible to artists and opens new avenues for creative expression, where the interaction between the participant and the artwork becomes more fluid and natural.

Moreover, the advent of embedding techniques like ResNet (short for Residual Network) \cite{he2016deep} and OpenAI's CLIP (Contrastive Language-Image Pre-training) \cite{radford2021learning} has added depth and sophistication to the interactions possible in interactive art. Embeddings represent unstructured, high-dimensional data—such as images, audio, or text—in numeric form within a lower-dimensional space, typically a matrix. This process allows for a more nuanced understanding of media, facilitating tasks such as image processing and extracting contextual information.

In the context of interactive art, these techniques enable a more detailed and sensitive interpretation of participants' actions and behaviors. For instance, face embeddings, which encode facial features into a compact numerical representation, can be employed in various sophisticated ways. Additionally, these embeddings can adeptly interpret different modalities like speech-to-text inputs or various camera inputs. When used in interactive art, such capabilities contribute to new forms of interaction.

As an example, CLIP can be used to create installations that respond to user input as text by finding semantically meaningful images through embeddings generated for both text and images. This strategy was employed in the interactive art installation "Dreampainter" (2021) by Varvara \& Mar \cite{canet2022dream}, which used CLIP to transform speech input from the audience into text and create interpretations of their dreams drawn by a robotic arm. As this example demonstrates, the technology enriches audience interaction and blurs the lines between the artwork and the audience, fostering a unique, co-creative experience where the artwork evolves in real-time in response to the viewer's input.

\begin{figure*}[h]
  \centering
  \includegraphics[width=\linewidth]{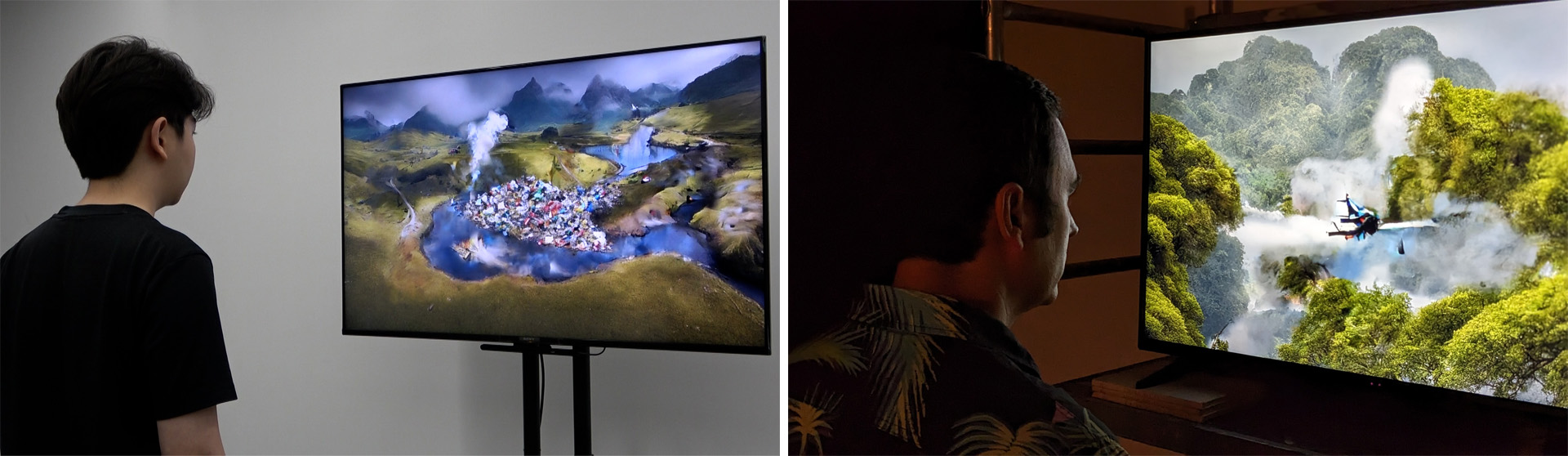}
  \caption{Audience interacting with the artwork using their gaze.}
  \Description{Audience interacting with the artwork.}
  \label{fig:artwork-interaction}
\end{figure*}

\subsection{Reference Artworks}

This research delves deeper into the realm of interactive art, particularly highlighting the advent of generative AI models in this field. A pivotal moment in this evolution was marked by the introduction of the GAN-based pix2pix model \cite{isola2017image} in 2017. Pix2pix showcased an innovative image-to-image translation interface, fundamentally transforming the landscape of interactive installations. It enabled real-time transformation of images, whether captured by a camera or drawn by users, into distinct and creative outputs. Despite being limited to lower resolutions and achieving only modest frame rates, primarily due to the GPU capabilities available then, pix2pix represented a significant breakthrough in the application of interactive AI \cite{akten2019learning}. This development demonstrated the potential of DL in artistic contexts while also opening new possibilities for artists and creators to explore real-time, AI-driven interactivity in their work.
A known example of using the pix2pix model in an interactive artwork is 'Learning to See' (2017) by Memo Akten \cite{akten2019learning}. To be more precise, this work can be described as a closed-circuit installation that "self-feeds and self-imagines" \cite{GeneYoungblood}: the real-time camera image composed of everyday objects manipulable by the audience is transformed into an abstract landscape by using generative AI.

Another relevant artwork is Circuit Training (2019) by Mario Klingemann \footnote{https://artsandculture.google.com/story/circuit-training-machine-made-art-for-the-people-barbican-centre/ngWRdP9M5scyLQ?hl=en}. This installation, based on GAN architecture, features a capture station, a screen that visualizes the dataset, and a curation system equipped with small touch screens. These screens are used to gather audience feedback on the outputs, enabling the system to learn and adapt to the style most favored by visitors. Participants wishing to be included in the training dataset of the piece must first give their consent by pressing a button on the touchscreen. This work is particularly fascinating as it unveils to the visitors the various components of AI architecture and exposes the often-unseen efforts required to construct AI systems, allowing spectators to engage as co-creators.

The "AI Portraits PRO" (2019) by Mauro Martino \footnote{https://www.mamartino.com/} in collaboration with Luca Stornaiuolo was an experimental interactive art net-art piece. Using a GAN trained with a dataset of 45,000 Renascence portraits from museum collections allowed users to transform themselves into classical art portraits through a website using a webcam \cite{schwab2019renaissance}. Since the algorithm was trained on Renaissance portraiture, which has its own stylistic and cultural biases, the work did not accurately represent modern or diverse facial features and expressions. For instance, smiling or laughing faces were not rendered correctly, reflecting the historical bias in portrait styles where such expressions were uncommon. Due to its immense popularity and the ensuing discussions about its biased outputs related to the dataset, the app was only available for a short span of a few days. This interactive piece holds significant relevance in demonstrating how an interactive system can illuminate new dimensions of meaning within a vast corpus of art through novel forms of inquiry and highlights the broader issue of bias in AI systems, where the characteristics of the training data heavily influence the output. Additionally, the work succeeded in actively engaging the public in a unique experience to explore, in this instance, the classical portraits housed in museums.

Dream Painter (2021) is an interactive robotic art installation by Varvara \& Mar that skillfully connects robotics and interface design to generative AI and lets the audience experience real-time navigation in the latent space of the model by sharing a dream in voice \cite{canet2022dream, guljajeva2022dream}. The sketch produced is interesting because it also allows the audience to complete the interpretation since it is between abstract and figurative, allowing a more open range of meanings from the person who interacted and also from the others. The authors analyzed the corpus of outputs from an exhibition in their study. They conducted an interesting classification of the results, focusing on the interactions between the participants and the algorithm. This analysis helped to understand how the models influenced the aesthetic qualities of the outputs \cite{guljajeva2023explaining}.

In addition, there are other interactive artworks that apply generative AI, such as 'Ray' (2021) by Zhang Weidi\footnote{https://www.zhangweidi.com/ray}, and 'Unsupervised' (2022) by Refik Anadol\footnote{https://www.moma.org/calendar/exhibitions/5535}. The latter does not demonstrate direct audience interaction but rather interaction with the system itself \cite{guljajeva2018interaction}.
As highlighted previously, the artworks that deploy co-creative generative AI tend to produce more meaningful results than closed systems \cite{guljajeva2021synthetic}. These cases, presented in this section, collectively illuminate the evolving relationship between AI technology and artistic expression, showcasing how AI can reshape artistic inquiry with audience interaction while also bringing to light the challenges and rising critical views on technology.

Since the presented artwork uses gaze as an interaction method, reviewing some historical interactive artworks that have also utilized gaze is essential. In 1992, the interactive artwork "De-Viewer"\footnote{Art+com, "De-Viewer," 1992. https://artcom.de/en/?project=de-viewer} by Art+com using eye-tracking presented at Ars Electronica marked a pivotal moment in this field \cite{sauter2008interfaces}, challenging the late 80s perception of computers merely as tools rather than creative mediums. The artists transitioned from traditional tools like brushes to digital interfaces such as the mouse, incorporating gaze to create a more interactive and engaging dialogue between the artwork and the viewer.

More contemporary examples include Golan Levin’s "Eye Code" (2007),\footnote{Golan Levin, "Eye Code," 2007, \url{https://www.flong.com/archive/projects/eyecode/index.html}.} which demonstrates the potential of eye-tracking in art by transforming the viewer's gaze into dynamic visual expressions. Anaisa Franco’s "Expanded Eye" (2008)\footnote{Anaisa Franco, "Expanded Eye," 2008, \url{https://www.anaisafranco.com/expandedeye}.} is an interactive light sculpture composed of a large transparent eye suspended from the ceiling. The sculpture recognizes the user's eye blinking and generates interactive animations based on it. Each blink multiplies the number of eyes in the projection in a fragmented, hexagonal, and dislocated way, mimicking the ultra-complex structure of insect compound eyes. Varvara \& Mar's "Binoculars to… Binoculars from…" (2013)\footnote{Varvara \& Mar, "BINOCULARS TO… BINOCULARS FROM…," 2013, \url{https://var-mar.info/binoculars/}.} explores the relationship between observation and interaction using gaze. This piece allows participants to see other places through a network of cameras in the city. Still, as a trade-off, their eye is projected into the place they are looking, creating a metaphor of the observer being observed.

These innovative works exemplify how eye-tracking and AI technology can deepen the interactive experience in media art, inviting viewers to engage in new ways. The presented work tries to merge gaze as a meaningful interaction method with the potential of AI generation to create unique experiential artwork.

\begin{figure*}[h]
  \centering
  \includegraphics[width=\linewidth]{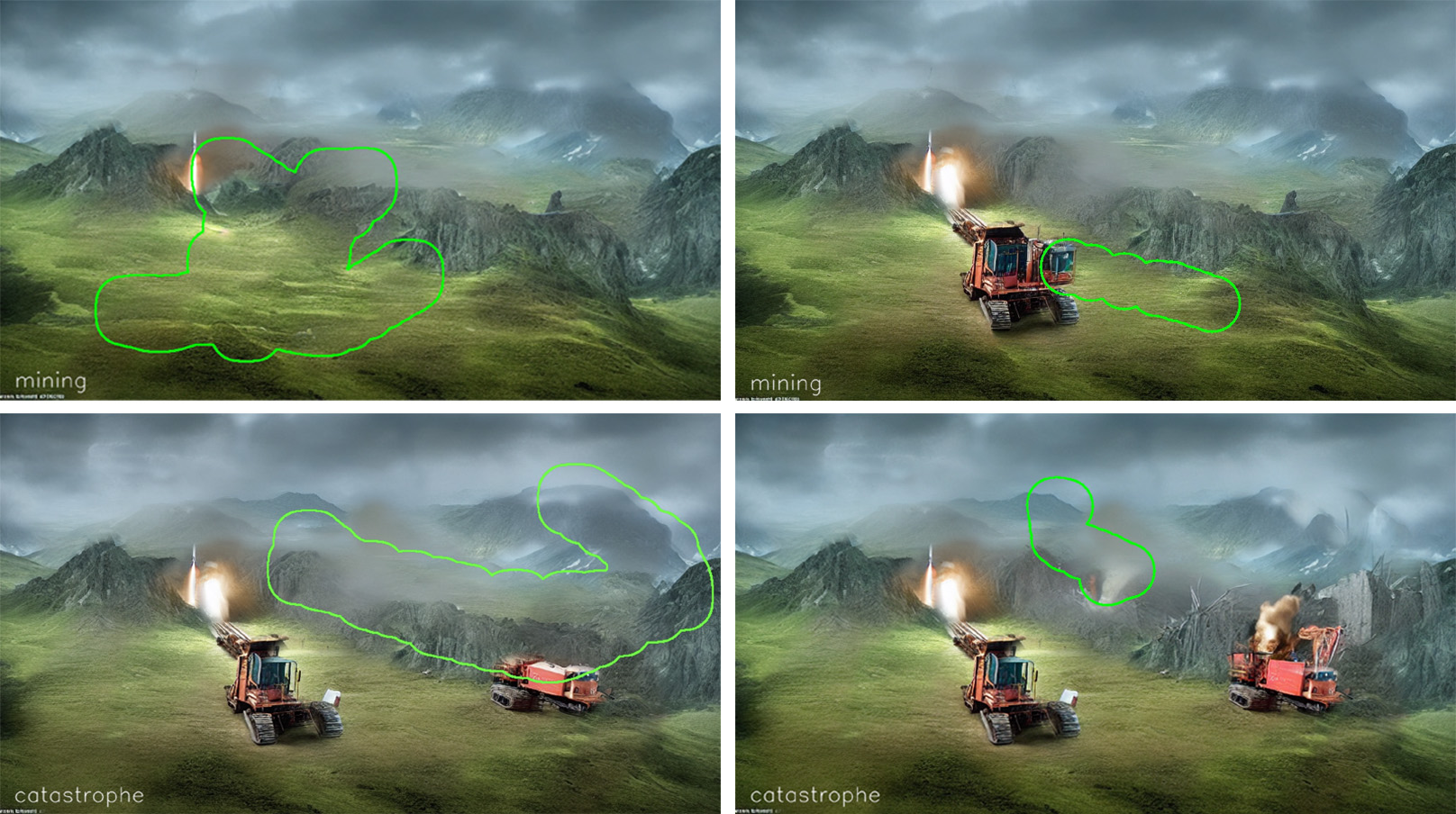}
  \caption{Debug view of the artwork showcasing the user's eye-tracking spots and a retrieved prompt. From left to right: prompt 'mining' affects the viewed area with the inpainting function.
}
  \Description{Debug view of the artwork displaying the user's eye-tracking spots and a retrieved prompt. From left to right: prompt 'mining' affects the viewed area with the inpainting function.
}
\label{fig:artwork-prompts}
\end{figure*}

\subsection{Historical and theoretical overview of the gaze}

The concept of "the gaze" has been a pivotal subject in critical, historical, and cultural theoretical discussions, particularly in the fields of art, literature, psychology, and human-computer interaction (HCI). Historically, the gaze has been explored extensively in the works of theorists such as Michel Foucault and Laura Mulvey. Foucault’s analysis of the gaze in "Discipline and Punish" highlights how visual observation serves as a means of exerting power and control \cite{foucault1977discipline}. Mulvey’s seminal essay "Visual Pleasure and Narrative Cinema" introduces the idea of the male gaze as a critique of commercial films, explaining how visual media often frames women as objects for male pleasure \cite{mulvey1975visual}. This theoretical framework underscores the power dynamics inherent in the act of looking and being looked at, revealing how the gaze can reinforce societal structures and gender roles.

Culturally, the gaze has been interpreted in various ways across different societies and eras. In Western art, for instance, the portrayal of subjects gazing directly at the viewer often served to establish a connection or assert dominance, while in Eastern art, the gaze might convey different symbolic meanings. These cultural interpretations of the gaze influence how viewers engage with and interpret visual works, making it a critical element in the creation and reception of art \cite{bryson1988gaze}.

From a scientific perspective, studying perception, vision, and gaze involves understanding the complex processes through which we interpret visual stimuli. Saccades, rapid movements of the eye that occur as we shift our focus from one point to another, play a crucial role in how we perceive our surroundings \cite{carpenter1988movements}. These eye movements are not random but are guided by our cognitive processes and prior knowledge \cite{findlay2003active}. Research in vision science reveals that saccades help gather detailed visual information, which the brain then processes to form a coherent understanding of the environment \cite{gilchrist2011saccades}.

In HCI domain, eye-tracking technology has become a vital tool for understanding user behavior and optimizing interface design. Eye-tracking studies provide insights into how users interact with digital interfaces, revealing patterns of attention and focus. For example, Jacob and Karn's comprehensive review on eye-tracking in HCI outlines its applications and the valuable data it provides for improving user experience \cite{jacob2003eye}. Duchowski’s extensive work on eye-tracking methodology and its applications in HCI further emphasizes its importance in understanding user interaction and enhancing the usability of interfaces \cite{duchowski2007eye}. By leveraging eye-tracking technology, researchers can design interfaces that align more closely with natural human visual behaviors, ultimately improving the efficiency and meaningfulness of user interactions \cite{carpenter2019towards}.

Integrating the critical, historical, and cultural discussions of the gaze with the scientific background of perception and vision, as well as the practical applications of eye-tracking in HCI, provides a comprehensive framework for understanding the impact of the gaze in visual art and digital interfaces. This approach not only highlights the theoretical significance of the gaze but also grounds it in empirical research and practical applications, offering a deeper insight into the viewer's experience and the thematic depth of the work.

\section{Vision of Destruction}

\begin{figure*}[h]
  \centering
  \includegraphics[width=\linewidth]{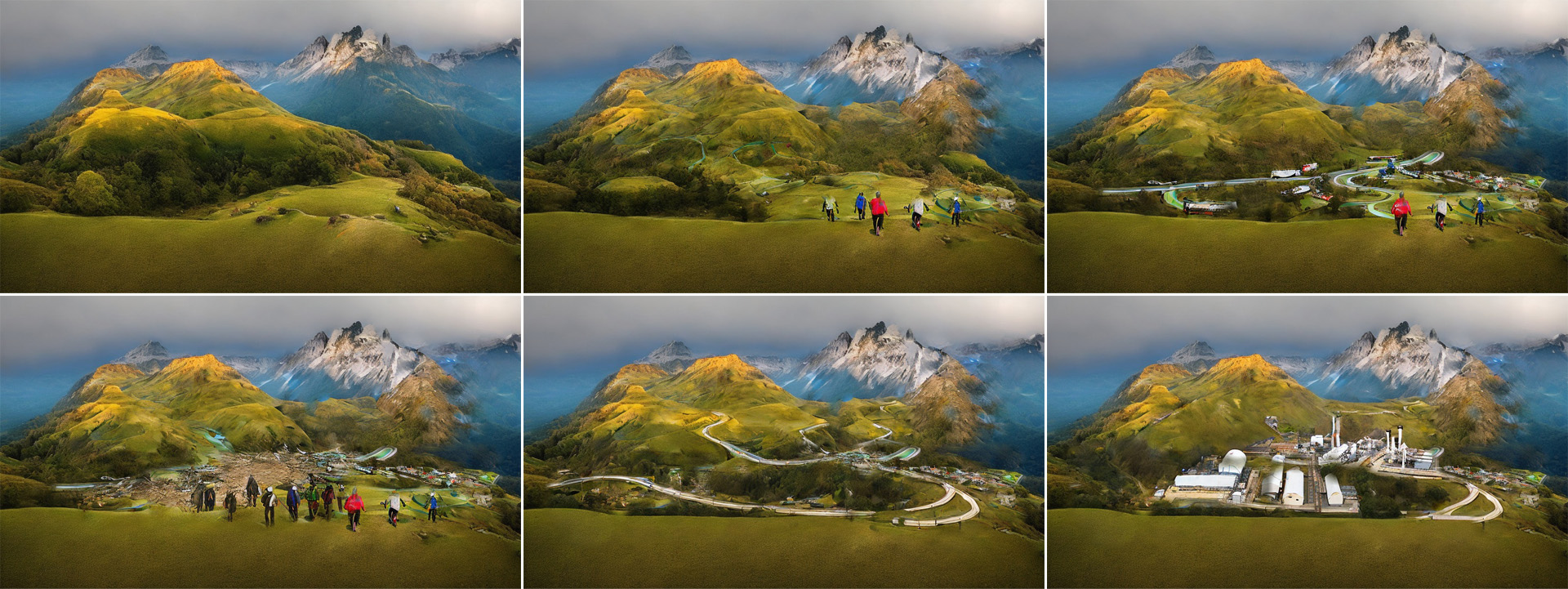}
  \caption{Progressive image transformation by the audience's gaze.}
  \Description{Progressive image transformation by the audience's gaze.}
  \label{fig:progressive-image}
\end{figure*}

"Visions of Destruction" is an immersive, interactive digital artwork that encapsulates human activities' transformative and often detrimental impact on Earth's natural landscapes \cite{10.1145/3623509.3635319}. This piece serves as a poignant commentary on the Anthropocene - the current geological age, viewed as the period during which human activity has dominated climate and the environment.
The artwork begins by displaying a series of breathtakingly beautiful natural landscapes, epitomizing the pristine and unblemished state of nature before significant human interference. These landscapes are vivid, dynamic representations of various ecosystems - from lush forests and tranquil lakes to majestic mountains and expansive deserts, all untouched by civilization.
As soon as a spectator engages with the artwork by directing their gaze towards it, the scenery begins to transform. This change is gradual yet inexorable, introducing artificial elements into the natural setting. What starts as a small path may slowly morph into a sprawling road and then into a bustling cityscape. Tranquil rivers may turn into industrial canals, and serene skies may become clouded with the smoke of factories. This metamorphosis is a powerful symbol of the degradation of nature due to industrialization and environmental catastrophe.

The transformation is driven by the audience's interaction. The direction of their gaze acts as a catalyst for change, making each experience unique. By using gaze-tracking technology, the artwork creates a dynamic and responsive environment where the spectator's attention directly influences the evolution of the landscape. This interactive element is pivotal in conveying the message that every action or inaction counts. It's a stark reminder that our collective choices have led to the current state of the environment. The motivation behind "Visions of Destruction" is to underscore the irreversible nature of climate change and to trace its origins to the seemingly minor steps of human progress.
The audience is not merely a passive observer but an active participant in this transformation. Through their engagement, they metaphorically move mountains, build cities, and launch rockets, feeling the weight of human impact on nature more profoundly. The artwork becomes a collage of generative AI, adapting and evolving based on the spectator's focus. This demonstrates the power of technology in art while symbolizing how technological advancements have shaped our environment.
In the end, "Visions of Destruction" aims to leave its audience with a lasting impression of their role in the Anthropocene. It's a call to action, a reminder of our planet's fragility, and an invitation to reflect on the consequences of human progress. This artwork is not just a display of the past and present but a forewarning of the future, urging viewers to consider the legacy they wish to leave for future generations.

\begin{figure}[htbp]
    \centering
    \begin{minipage}{0.45\textwidth}
        \centering
        \includegraphics[width=\linewidth]{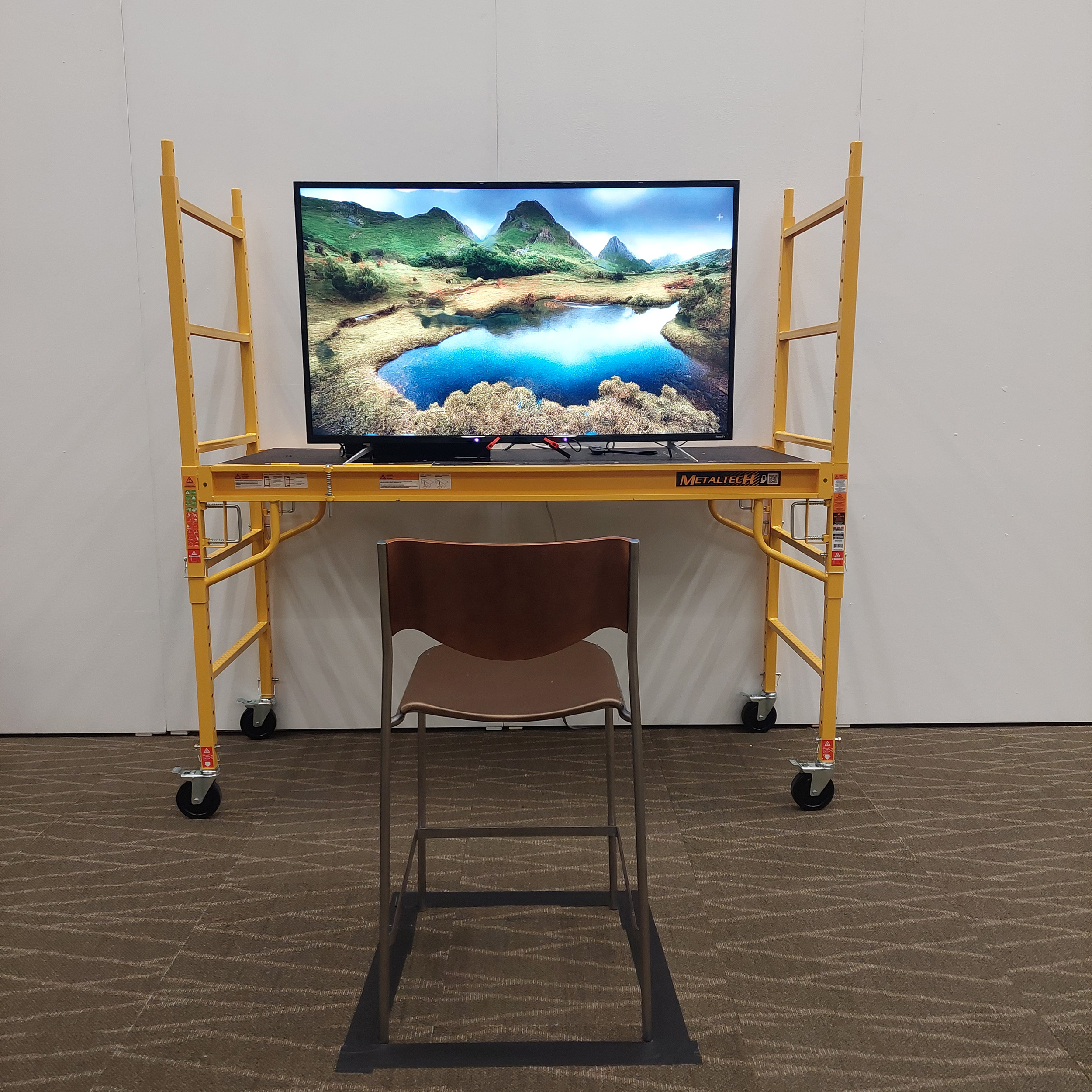}
        \caption{Installation setup in the AI Gallery at the CVPR Conference 2024 in Seattle, US, curated by Luba Elliot}
        \label{fig:setupInstallation}
    \end{minipage}\hfill
    \begin{minipage}{0.45\textwidth}
        \centering
         \includegraphics[width=\linewidth,page=1]{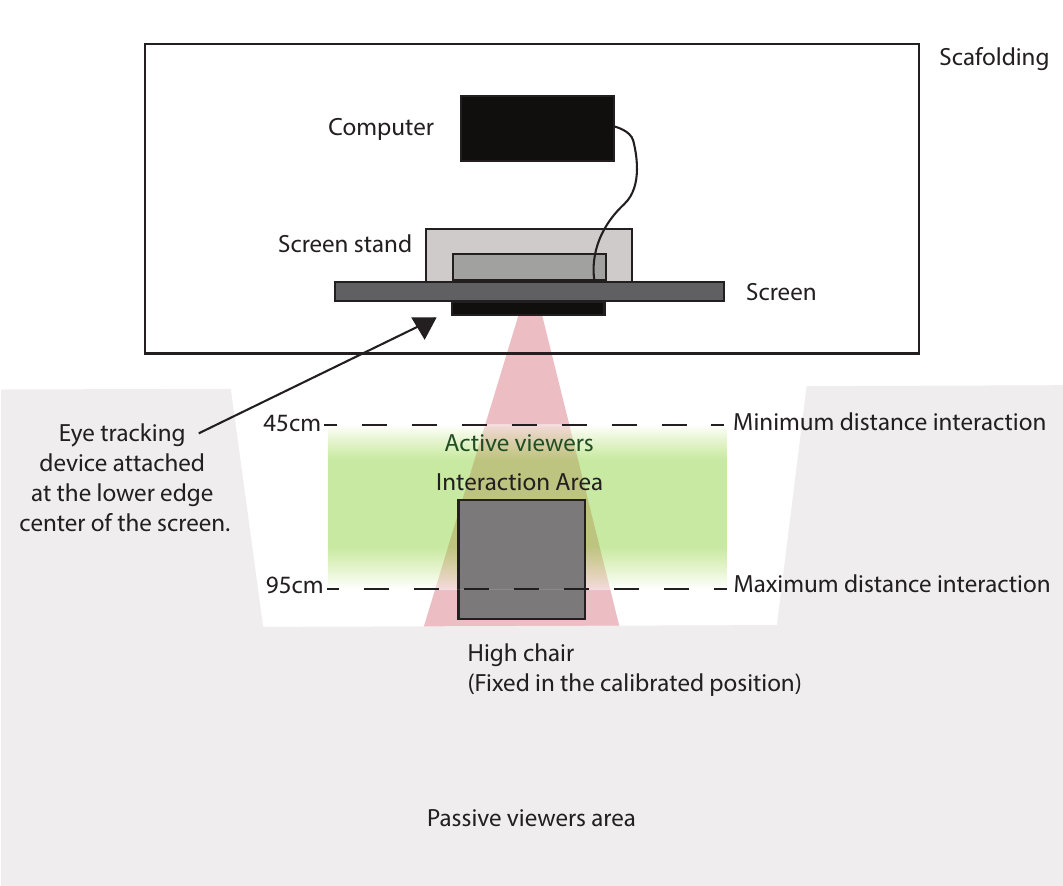}
        \caption{Floor Plan  of the art installation}
        \label{fig:floorplan}
    \end{minipage}
    
\end{figure}

\subsection{Method}

"Visions of Destruction" can be seen as a collage of generative AI. The spectator's gaze indicates where a change will appear on a digital canvas. The regions of gaze focus are constructed using aggregated spectator gaze data collected through high-resolution eye-tracking technology. For this purpose, we are utilizing the Tobii Eye Tracker 5 device \footnote{https://gaming.tobii.com/product/eye-tracker-5/}, specifically designed for gaming and interactive experiences. The device operates effectively without requiring individual calibration for each user, making it highly suitable for art installations. The primary consideration is that users must be close to the screen and within its eye-tracking range. Due to varying heights among spectators, the system functions optimally when the audience is seated, reducing height differences by nearly half. Additionally, the use of adjustable chairs further enhances convenience and effectiveness. Eye-tracking is not affected by the room's light conditions since it operates using infrared light. The sensor is installed at the lower part of the screen. For a better understanding of the installation setup and its involved components, see Figures \ref{fig:setupInstallation}, \ref{fig:floorplan}.

It's noteworthy to note that glasses can interfere with the eye-tracking system, particularly when reflections occur. However, even with glasses, the system generally functions well, though there may be brief moments where tracking is momentarily lost. In this installation, if there are no spectators for a few seconds, the landscape automatically regenerates into a new one. The regeneration process in the installation serves as a metaphor, symbolizing how nature, given sufficient time, can also recover and heal from the damages inflicted upon it.

\begin{figure}[h]
  \centering
  \includegraphics[width=\linewidth]{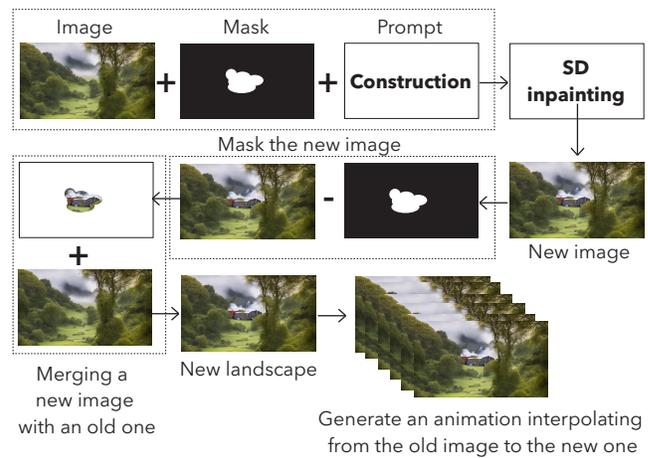}
  \caption{Image generation process using custom software based in Stable Diffusion.}
  \Description{Schema of the image generation process using custom software based in Stable Diffusion.}
  \label{fig:artwork-technical-schema}
\end{figure}

The artwork begins with an AI-created depiction of idyllic beautiful landscapes generated through Stable Diffusion \cite{rombach2022high}, offering a romanticized view of nature based on collective human memories and trained on the LAION-5B dataset \cite{10.1145/3623509.3635319}. In a technical twist, the data from the eye-tracker is used to craft a mask that alters specific areas of the image where the audience's gaze lingers (see Figure \ref{fig:artwork-interaction}). This involves applying an array of artist-designed prompts about the climate emergency to perform inpainting with Stable Diffusion, such as “mining” or “catastrophe” (see Figure \ref{fig:artwork-prompts}). The result is a dynamic experience for viewers witnessing a seamless morphing of nature as if the gaze is a weapon of mass destruction.

The system employs a recursive approach, periodically iterating over eye-tracking data to create evolving landscapes. This process incorporates aesthetic accidents as part of the piece's aesthetics, as some prompts yield unexpected outcomes. Initially, the landscapes undergo slower transformations due to drastic changes requested by the prompts, leading to more rapid changes with continued interaction (Figure \ref{fig:progressive-image}). Technical challenges arose during development, especially related to image degradation occurring with repeated inpainting. The solution involved using a masked area for inpainting and merging only this part with the original image, leaving the rest unchanged (see Figure \ref{fig:artwork-technical-schema}). The output is an animation generated using PyTorch's version of the AI model 'FILM: Frame Interpolation for Large Motion' \cite{reda2022film}, which interpolates between the old and new images. This installation capitalizes on recent advancements in fast image generation using diffusion models, a key aspect of interactive AI art that requires computation with minimal delays.

Through practice-based research, we explore how generative AI could serve the purpose of interaction in art. For delivering real-time experiences to the audience, we combine gaze-based interaction by using eye-tracking technology and artists' created prompts that serve the artistic concept. When comparing this methodology to the GAN era, then we see drastic change: previously, artists were composing small datasets \cite{vigliensoni2022small}, making network bending to create errors \cite{broad2021network} and curating GAN's output \cite{guljajeva2023artistic}. With diffusion models like Stable Diffusion, while curating outputs remains common when these models are not used for interactive art, artists have shifted their main practices to prompt-guided creations.

\section{DISCUSSION}

In interactive AI art, the artist's role is pivotal in shaping the audience's experience, much like in other interactive art forms. However, a unique aspect of generative AI art is its inherent unpredictability. Artists must adapt to and incorporate the spontaneous nature of AI-generated content, viewing these unexpected outcomes not as limitations but as integral, dynamic components of artistic expression. This approach marks a distinct shift in art creation and experience, blending creativity with the unforeseen capabilities of AI to generate unique and individualized content.

Interactive artworks utilizing AI require open-source models due to their real-time nature and the need for custom software, ideally running on a local machine. For instance, this project runs on a gaming machine with an NVIDIA RTX 4090. While APIs could be a viable alternative in some scenarios, they are subject to network lag times, usage fees, and the risks associated with the provider's changing commercial strategies, potentially leading to closure or restricted access. API request fees can lead to substantial costs, particularly in exhibitions with high visitor traffic, making it challenging for artists to manage, as we have observed in our experience with art institutions. From a preservation standpoint, it is more advantageous to have a system running entirely on one or more local machines using open-source libraries, ensuring long-term sustainability. Rafael Lozano-Hemmer highlights in his guide to the preservation of media art both the significance and the risks associated with using closed-source products that may eventually be discontinued, necessitating the re-engineering of artworks with open-source tools \cite{lozano2019best}.

\section{CONCLUSIONS}
In conclusion, the study effectively demonstrates the transformative power of generative AI in interactive art. Through the innovative use of gaze-tracking technology as an interface for AI, the artwork vividly symbolizes the environmental impacts of human activities. This research highlights the shift from traditional artistic methods to prompt-guided creation using diffusion models like Stable Diffusion, marking a significant evolution in interactive art. Furthermore, it underscores the importance of open-source models for real-time interaction and sustainability in art installations. The study contributes to a deeper understanding of AI's role in artistic expression and audience engagement, bridging technology and environmental awareness in a profound and impactful way.

\begin{acks}
Mar Canet Sola and Varvara Guljajeva are the authors of the idea and conceptualization of 'Visions of Destruction'. Mar Canet Sola and Isaac Clarke did the software's technical development. Mar Canet Sola is supported as a CUDAN research fellow and ERA Chair for Cultural Data Analytics, funded through the European Union's Horizon 2020 research and innovation program (Grant No. 810961).

\end{acks}

\bibliographystyle{ACM-Reference-Format}
\bibliography{sample-base}

\appendix

\end{document}